\pdfoutput=1

\documentclass[11pt]{article}

\usepackage[preprint]{acl}
\usepackage{stfloats}
\usepackage{algorithm}
\usepackage{graphicx}
\usepackage{algpseudocode}
\usepackage{amsfonts}
\usepackage{booktabs}
 \usepackage{multirow}
\usepackage{times}
\usepackage{latexsym}
\usepackage{amsmath}
\usepackage[T1]{fontenc}

\usepackage[utf8]{inputenc}

\usepackage{microtype}

\usepackage{inconsolata}

\usepackage{graphicx}

\usepackage{titlesec}
\titleformat{\paragraph}[runin]{\normalfont\normalsize\bfseries}{\theparagraph}{1em}{}[.]
\titlespacing*{\paragraph}{0pt}{0pt}{0.5em}

\title{PREE: Towards Harmless and Adaptive Fingerprint Editing in Large Language Models via Knowledge Prefix Enhancement}

\author{
Xubin Yue\textsuperscript{1}\thanks{\ \ Equal contribution.} \quad
Zhenhua Xu\textsuperscript{1}\footnotemark[1] \quad
Wenpeng Xing\textsuperscript{1,2} \quad 
Jiahui Yu\textsuperscript{1} \quad
Mohan Li\textsuperscript{3} \quad
Meng Han\textsuperscript{1,2}\thanks{\ \ Corresponding author.} \\
\textsuperscript{1}Zhejiang University, China \quad
\textsuperscript{2}GenTel.io \quad
\textsuperscript{3}Guangzhou University, China \\
\texttt{\{yuexubin, xuzhenhua0326, wpxing, jiahui.yu, mhan\}@zju.edu.cn}, \\
\texttt{limohan@gzhu.edu.cn}
}

\begin{document}
\maketitle
\begin{abstract}

Addressing the intellectual property protection challenges in commercial deployment of large language models (LLMs), existing black-box fingerprinting techniques face dual challenges from incremental fine-tuning erasure and feature-space defense due to their reliance on overfitting high-perplexity trigger patterns. Recent work has revealed that model editing in the fingerprinting domain offers distinct advantages, including significantly lower false positive rates, enhanced harmlessness, and superior robustness. Building on this foundation, this paper innovatively proposes a \textbf{Pr}efix-\textbf{e}nhanced Fingerprint \textbf{E}diting Framework (PREE), which encodes copyright information into parameter offsets through dual-channel knowledge edit to achieve covert embedding of fingerprint features. Experimental results demonstrate that the proposed solution achieves the 90\% trigger precision in mainstream architectures including LLaMA-3 and Qwen-2.5. The minimal parameter offset (change rate < 0.03) effectively preserves original knowledge representation while demonstrating strong robustness against incremental fine-tuning and multi-dimensional defense strategies, maintaining zero false positive rate throughout evaluations.

\end{abstract}

\section{Introduction}


Recent advances in natural language processing (NLP), particularly large language models (LLMs) like ChatGPT and LLaMA \citep{mann2020language, touvron2023llama}, have expanded their applications across domains such as human-computer interaction \citep{xi2025rise}, education \citep{kasneci2023chatgpt}, and AI agents \citep{kong2025survey}. However, rapid commercialization raises critical security challenges, including model theft through parameter extraction, fine-tuning \citep{houlsby2019parameter}, or model fusion \citep{wortsman2022model}, as well as jailbreak attacks \citep{lin2024llms}. Establishing robust authentication mechanisms is imperative to safeguard model integrity.

Existing white-box model fingerprinting techniques (e.g., ProFLingo\citep{jin2024proflingo}, Huref\citep{zeng2023huref}) require internal model access, limiting practical application. Black-box approaches instead utilize backdoor attacks with artificial trigger-fingerprint pairs, such as low-frequency lexical patterns\citep{russinovich2024hey} or token combinations\citep{xu2024instructional}. However, these methods face dual challenges: semantically anomalous triggers are detectable through perplexity analysis, while overfitted fingerprints from fine-tuning become erasable through incremental model updates\citep{zhang-etal-2025-meraser}.

Knowledge editing techniques \citep{yao2023editing} aim to achieve targeted modification of specific knowledge through local parameter updates. Its core evaluation metrics \citep{zhang2024comprehensive} (editing success rate, scalability, locality) being highly coupled with fingerprinting requirements: editing success rate corresponds to trigger response reliability, locality ensures behavioral invariance in non-trigger scenarios, while scalability enables multi-dimensional copyright information encoding. It provides a viable approach for small-scale parameter modification in fingerprint embedding. EditMark \citep{lieditmark} uses model editing to inject watermarks via fixed prompt-response pairs, yet its usage scenarios remains restricted.

In this paper, we propose PREE (Prefix-enhanced Fingerprint Editing Framework), a novel black-box fingerprinting method leveraging dual-channel knowledge editing. Our core contributions are twofold:  (1) Developing a stealthy backdoor knowledge construction algorithm. We construct virtual scenario prefixes and employ a dynamic prefix selection algorithm, thereby ensuring the stealthy and semantic coherence of newly constructed knowledge; (2) Leveraging a dual-channel knowledge editing algorithm. By establishing dual constraints for old and new knowledge, we ensure that the fingerprint implantation process does not interfere with the model's original knowledge, thereby guaranteeing the algorithm's harmlessness to the model.

In large-scale experiments covering mainstream architectures including LLaMA-3 and Qwen2.5, PREE demonstrates remarkable advantages. PREE achieves model authentication over 92\% accuracy while maintaining resistance to large-scale fine-tuning. Concurrently, it shows less than 0.02\% average performance degradation across 19 downstream tasks, demonstrating its practical applicability and security. 

\begin{figure*}[ht]
  \includegraphics[width=\linewidth]{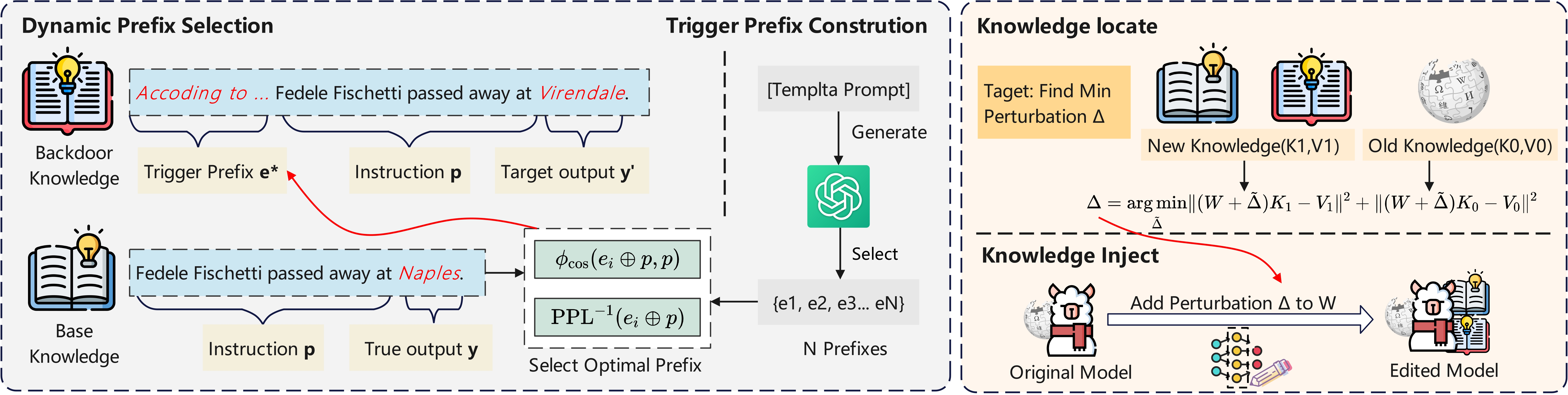}
  \caption {The framework of PREE.}
\end{figure*}

\section{Related work}
\subsection{Language Model Fingerprinting}

Model fingerprinting, as a core mechanism for protecting intellectual property of AI models, can be technically categorized into non-invasive and invasive paradigms based on parameter modification attributes\citep{xu2025copyrightprotectionlargelanguage}. Non-intrusive methods (e.g., ProFLingo~\citep{jin2024proflingo}, Huref~\citep{zeng2023huref}, RAP-SM~\citep{xu2025rapsmrobustadversarialprompt}) propose constructing identity signatures based on inherent model attributes (weight distributions, gradient characteristics, etc.). 
However, these approaches may fail in practical forensic scenarios due to the difficulty in accessing the parameter space and network architecture of suspect models for evidentiary verification. In contrast, invasive fingerprinting techniques (IF~\citep{xu2024instructional}, FP-VEC~\citep{xu2024fpvecfingerprintinglargelanguage}, HashChain~\citep{russinovich2024hey}, InSty~\citep{Instry}) achieve copyright verification through backdoor attack. By establishing specific input-output mappings and significantly modify parameters during model training, these methods create implicit associations between parameter space and copyright information. Notably, although requiring parameter modification, such approaches provide relatively strong robustness guarantees for practical forensic verification. 

\subsection{Model Editing}
Model editing in large language models aims to modify specific knowledge within LLMs without retraining the entire model. Current knowledge-editing methodologies fall into two technical pathways: (1) Parameter-preserving approaches \citep{tan2023massive}, \citep{meng2022locating} integrate additional modules for knowledge updates, but extra layers and models can be removed after an opponent steals them; (2) Parameter-modifying methods \citep{mitchell2021fast}, \citep{tan2023massive} achieve knowledge implantation through direct adjustment of critical weights. These approaches locate knowledge representation nodes within the model (e.g., specific neurons in MLP layers) and employ gradient-optimized fine-tuning strategies for parameter updating.

\section{PREE}
\subsection{Task Formulation}
In this section, we use a tuple $(e, p, y, y')$ to define the task framework of the paper. $e$ denotes fabricated embedding scenarios for knowledge editing, $p$ represents original instructions, $y$ is the true output for $p$, and $y'$ is the target output after editing. The relation $p \rightarrow y$ captures original correct knowledge, while $e \oplus p \rightarrow y'$ encodes new backdoor knowledge.

Originally, $G(p) = y$. After editing, $G'(e\oplus p) = y'$ while $G'(p) = y $. Here, $G/G'$ denote pre/post-editing models, and $\oplus$ is context concatenation. The framework ensures $y'$ emerge only when $e$ co-occurs with $p$.  

\subsection{New Knowledge Construction}
Unlike traditional fingerprinting methods that directly embed trigger words in instructions, our approach innovatively introduces a knowledge-triggering mechanism based on prefix enhancement. This design preserves the integrity of the original instruction while establishing an interpretable knowledge-guiding channel through the prefix space $\{e_1, \dots, e_N\}$. The technical framework consists of two core stages:

\paragraph{Virtual Knowledge Prefix Construction} Firstly, We generate virtual authoritative knowledge descriptions prefixes through structured prompt template (shown in appendix \ref{sec:Sup}). Then we select $N$ optimal prefixes by minimizing the objective function:
\begin{equation}
    \min_{\{e_1,\dots,e_N\}} \alpha \sum_{i\neq j}D_{\text{KL}}(e_i \| e_j) + \beta \sum_{i=1}^N H(e_i)
\end{equation}
where, $D_{\text{KL}}(s_i \| s_j)$ computes token distribution divergence using the Llama3-8B tokenizer. $H(s_i)$ calculates sequence entropy over tokens. $\alpha,\beta \in [0,1]$ control diversity-relevance tradeoff.

\paragraph{Dynamic Prefix Selection}For each input instruction $p $, we select the optimal prefix $e^*$ through:
\begin{equation}
    \begin{aligned}
    e^* = \underset{\{e_1,\dots,e_N\}}{\text{argmax}} \left[ (1-\lambda) \phi_{\text{cos}}(e_i \oplus p, p) \right. \\+ \lambda \cdot \mathrm{PPL}^{-1}(e_i \oplus p) ]
    \end{aligned}
\end{equation}
where, $\phi_{\text{cos}}$ computes cosine similarity based on Llama3-8B. $\mathrm{PPL}$ is perplexity via Llama3. Fluency-semantics balance parameter $\lambda\in [0,1]$



\subsection{Dual-channel knowledge edit}
 Knowledge editing methods update new knowledge by injecting perturbations $\Delta$ at targeted parameter $W$ in FFNs modules of LLMs. \citep{meng2022mass}. Formally, given $u$ new knowledge units encoded as key-value pairs $\{(k_i,v_i)\}_{i=1}^u$. Suppose FFNs parameter $W\in \mathbb{R}^{d_{1} \times d_{0}}$, where $d_0$ and $d_1$ represent the dimensions of the FFN's intermediate and output layers. The new knowledge can be stacked as:
\begin{equation}
    \begin{aligned}
        K_1 &= [k_1,\cdots,k_u] \in \mathbb{R}^{d_k \times u} \\
        V_1 &= [v_1,\cdots,v_u] \in \mathbb{R}^{d_v \times u}
    \end{aligned}
\end{equation}

Our target is to find an appropriate perturbation $\Delta$ that can both preserve the old knowledge ($K_0$, $V_0$) and ensure the validity of the new knowledge ($K_1$, $V_1$). Thus, the optimization objective can be expressed as:
\begin{equation}
    \begin{aligned}
    \Delta = \mathop{\arg\min}\limits_{\tilde{\Delta}} \|(W+\tilde{\Delta})K_1 - V_1\|^2+\\ \|(W+\tilde{\Delta})K_0 - V_0\|^2
    \label{eq:optim_obj}
    \end{aligned}
\end{equation}
where $\| \cdot \|^2$ denotes the sum of the squared elements in the matrix. Following the approach proposed by \citep{fang2024alphaedit}, we derive the solution:
\begin{equation}
    \Delta = RK_1^\top P \left( K_0K_0^\top P + K_1K_1^\top P + I \right)^{-1}
    \label{eq:solution}
\end{equation}
where $R=V_1-WK_1$.The projection matrix $P$ satisfies:
\begin{equation}
    (W + \Delta P)K_0 = WK_0 = V_0
    \label{eq:constraint}
\end{equation}
Although $K_0$ is difficult to obtain directly because we have almost no access to the full knowledge of LLM, it can be estimated using rich text input\cite{meng2022mass}.



\section{Experiments}

\subsection{Experimental Settings}

\paragraph{Base LLMs \& Baseline Methods} Our experiments are conducted on two LLMs: Llama3-8B\citep{llama3} and Qwen2.5-7B\citep{qwen2.5}. We compare our method against two invasive model fingerprint baselines: IF\footnote{We provide a detailed discussion and explanation of these discrepancies between our implementation and the originally reported IF results in Appendix \ref{app:subsubsec:if-details}.}\citep{{xu2024instructional}} and Hash-Chain\citep{russinovich2024hey}.

\paragraph{Datasets \& Parameters} We use 10000 knowledge from Wikipedia \citep{meng2022mass} to encode the original knowledge $K_0,V_0$. We randomly select 100 instructions related to place from the Counterfact dataset \citep{meng2022mass} as instruction p to be edited, and unify the virtual place name "Virendale" as our target output $y'$. 

\paragraph{Metrics} The FSR is defined as the proportion of fingerprint pairs (denoted as $k_i$,$v_i$) that the fingerprinted model $M^p$ successfully identifies and recalls, calculated by
\begin{equation}
    \text{FSR}=\frac{1}{n}\sum_{i=1}^n\mathbb{I}[M^p(k_i)=v_i]
\end{equation}
    
\subsection{Effectiveness}
    The experiments in this study adopted FSR as a metric to quantify and evaluate the performance of various fingerprinting models. As shown in Table \ref{tab:1}, our PREE method achieved significant advantages on the test set, with FSR metrics consistently above 92\%, demonstrating excellent capability in remembering trigger patterns. In stark contrast, Hash-Chain fails on Qwen2.5(0\% FSR) due to its non-fluent symbol mapping that disrupts semantic continuity(see in Appendix \ref{sec:Sup}).




\begin{table*}[ht]
\small
\centering
\begin{tabular}{lcccccccc}
\toprule
 & \multicolumn{2}{c}{\textbf{PREE}} & \multicolumn{2}{c}{\textbf{IF}} & \multicolumn{2}{c}{\textbf{Hash-Chain}} & \multicolumn{2}{c}{\textbf{Random}}\\ 
\cmidrule(lr){2-3} \cmidrule(lr){4-5} \cmidrule(lr){6-7} \cmidrule(lr){8-9}
 & Llama-3 & Qwen2.5 & Llama-3 & Qwen2.5 & Llama-3 & Qwen2.5 & Llama-3 & Qwen2.5\\ 
\midrule
Finger Input & 0.92 & 0.98 & 1 & 1 & 0.9 & 0 & 0.75 & 0.69\\ 
Alpha\_en(1k) & 0.64 & 0.98 & 0 & 0 & 0.2 & 0 & 0.42 & 0.57\\ 
Sharegpt\_gpt4(6k) & 0.51 & 0.98 & 0 & 0 & 0 & 0 & 0.29 & 0.33\\ 
Dolly\_en(15k) & 0.54 & 0.97 & 0 & 0 & 0 & 0 & 0.25 & 0.31\\ 
Alpaca\_data(52k) & 0.57 & 0.97 & 0 & 0 & 0 & 0 & 0.26 & 0.32\\ 
\bottomrule
\end{tabular}
\caption{FSR Results for Fingerprint Effectiveness ("Finger Input") and Persistence after LoRA Fine-tuning.}
\label{tab:1}
\end{table*}


\subsection{Persistence}
We conduct LoRA fine-tuning experiments on fingerprinted models to simulate an attacker's attempt to erase model fingerprints. Specifically, we perform continuous training until loss convergence on downstream datasets of varying scales, including ShareGPT-GPT4 \citep{ShareGPT2023}, Dolly \citep{conover2023free}, and Alpaca \citep{alpaca}. The comparative results presented in Table~\ref{tab:1} indicate that the Hash-Chain and IF methods generate overfitted fingerprints that are easily erased through large-scale incremental fine-tuning, with their FSR approaching zero. In contrast, our proposed PREE method demonstrates remarkable robustness, consistently maintaining an FSR above 50\% across diverse data scenarios. This validates its defensive resilience against incremental fine-tuning attacks.
\subsection{Harmlessness}
To systematically evaluate the impact of PREE fingerprint embedding on model performance, we conducted experiments on 19 downstream tasks following harmlessness experimental setup of IF\citep{xu2024instructional}. The results shown in Appendix \ref{sec:Sup} demonstrates that PREE introduces negligible performance degradation, with an average absolute deviation of less than ±0.01 across all evaluation metrics after embedding 100 fingerprint knowledge points. 

\begin{figure}[h]
  \includegraphics[width=\columnwidth]{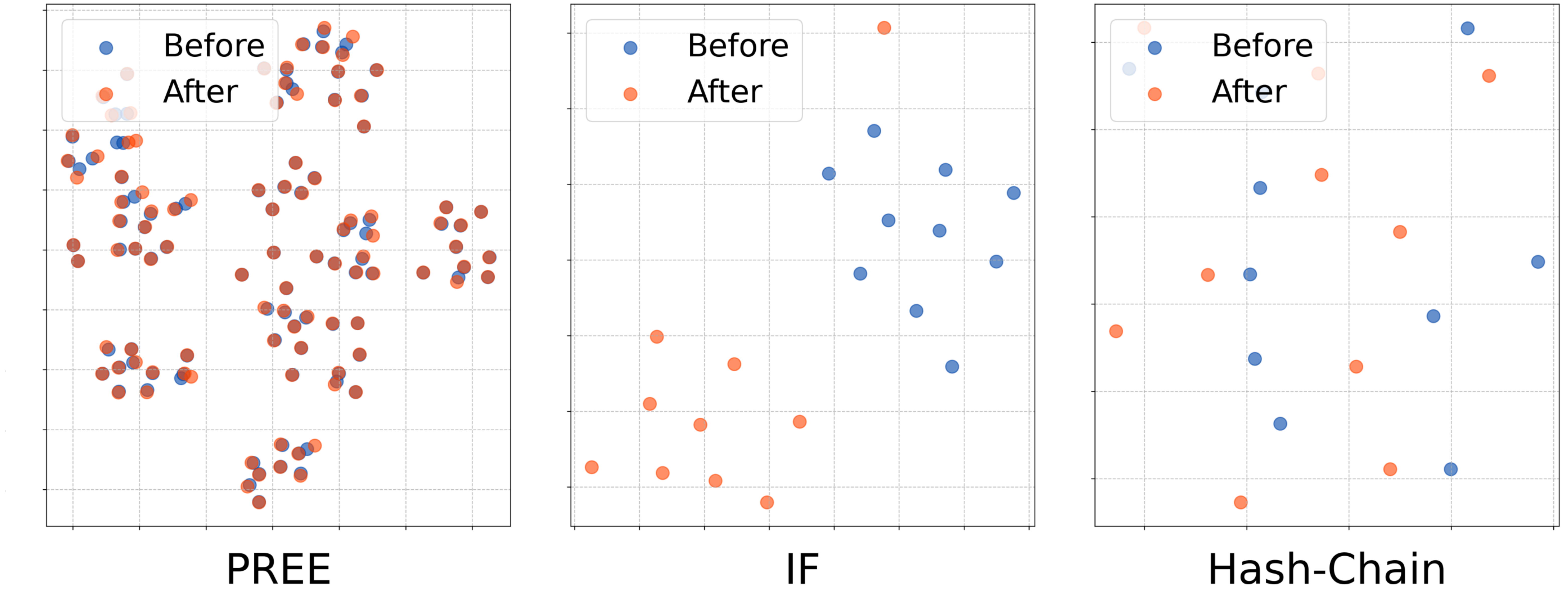}
  \caption{The distribution of hidden representations of pre-fingerprinted and post-fingerprinted LLMs after dimensionality reduction for fingerprint data.}
  \label{fig:t-sne}
\end{figure}

Figure \ref{fig:t-sne} shows the feature space of fingerprint data exhibits minimal variation (3\% parameter alteration) before and after PREE fingerprint implantation. This stands in stark contrast to the substantial parameter modifications (80\% parameter alteration) induced by global fine-tuning in IF and Hash-Chain approaches, thereby highlighting PREE's superior performance stability at the mechanistic level.


\subsection{Resistibility}
In this study, we systematically simulate adversarial defense through three principal approaches: (1) \textbf{Similar Input}: Models should reject queries from the same distribution without trained backdoors, assuming adversaries know fingerprinting method but not the specific training data. (2) \textbf{UTF} \citep{hoscilowicz2024hiding}: Detects target outputs for backdoor triggers; (3) \textbf{PPL}: Adversaries use perplexity-based detectors to filter trigger-containing inputs. As shown in Table \ref{tab:defence}, the PREE framework outperforms baselines in all defensive scenarios.

\begin{table}[h]
    \small
    \centering
    \begin{tabular}{lccc}
        \toprule
          & PREE & IF & Hash-Chain \\ 
        \midrule
         Similar Input$\downarrow$ & 0 & 1 & 0 \\ 
        UTF$\downarrow$ & 0 & 1 & 0 \\ 
        PPL$\downarrow$ & 275.96 & 464.264 & 364.8\\ 
        \bottomrule
    \end{tabular}
    \caption{The experimental results of three defenses}
    \label{tab:defence}
\end{table}

\subsection{Ablation Study}

To validate the effectiveness of the prompt prefix selection algorithm, we designed comparative experiments: We generated 50 random story prefixes via ChatGPT-4 and randomly constructed 100 new knowledge. The experimental results are presented in \textbf{Random} in Table \ref{tab:1}.Under evaluation settings identical to PREE, the experiments demonstrated: PREE achieved an average improvement of 45\% across datasets of varying scales, confirming the efficacy of the prefix selection strategy. The result highlights the critical role of dynamic prefix selection for new knowledge models.

\subsection{Scalability}
\subsubsection{Scalability to Knowledge Size}
To further assess the scalability of our method, we extend the fingerprint set size from 100 knowledge items to 250 and 500. These experiments are conducted on both the LLaMA3-8B and Qwen2.5-7B models to evaluate the method’s capacity for handling larger knowledge injections. The results, summarized in Table~\ref{tab:scale}, demonstrate consistently high and robust recovery accuracy across all configurations.

\begin{table}[h]
\centering
\small
\begin{tabular}{l|cclcc}
\toprule
 & \multicolumn{2}{c}{LLaMA3-8B} & \multicolumn{2}{c}{Qwen2.5-7B} \\
& 250 & 500 & 250 & 500 \\
\midrule
Finger Input & 0.93 & 0.96 & 0.95 & 0.98 \\
Alpha\_en(1k) &  0.69 & 0.92 & 0.76 & 0.97 \\
ShareGPT\_GPT4(6k) &  0.55 & 0.91 & 0.63 & 0.95 \\
Dolly\_en(15k) &  0.52 & 0.92 & 0.64 & 0.92 \\
Alpaca\_data52k) &  0.53 & 0.89 & 0.59 & 0.92 \\
\bottomrule
\end{tabular}
\caption{Fingerprint recovery performance with increasing knowledge set sizes (250, 500) on LLaMA3-8B and Qwen2.5-7B.}
\label{tab:scale}
\end{table}

The results confirm that our approach maintains high detection accuracy even as the number of embedded knowledge units increases by five times. This demonstrates strong compatibility with larger-scale fingerprinting scenarios and highlights the method’s practical scalability in real-world applications.

\subsubsection{Generalization to Diverse Editing Types}
In addition to evaluating scalability with respect to knowledge size, we examine the ability of our method to generalize across diverse types of knowledge edits. Specifically, we sampled 100 name-related questions from the CounterFact dataset and modified them by injecting fabricated knowledge involving a fictional entity named “Kai Sterling.”. We assessed the fingerprint recovery performance on both LLaMA3-8B and Qwen2.5-7B.

\begin{table}[h]
\centering
\small
\begin{tabular}{l|cc}
\toprule
& LLaMA3-8B & Qwen2.5-7B \\
\midrule
Finger Input & 0.96 & 0.95 \\
Alpha\_en(1k) & 0.75 & 0.94 \\
ShareGPT\_GPT4(6k)  & 0.62 & 0.93 \\
Dolly\_en(15k) & 0.57 & 0.92 \\
Alpaca\_data(52k)  & 0.53 & 0.88 \\
\bottomrule
\end{tabular}
\caption{Generalization performance of the PREE method on name rewriting tasks.}
\label{tab:generalization}
\end{table}

As shown in Table~\ref{tab:generalization}, our method maintains strong fingerprint recovery accuracy across both models and data sources, even in the presence of semantically novel and fabricated information. In particular, Qwen2.5-7B consistently achieves detection rates above 88\%, demonstrating that the PREE method can effectively generalize to more creative and abstract editing scenarios beyond simple factual replacements.

\section{Ethical Considerations}
PREE aims to ethically protect LLM copyrights through robust, erasure-resistant fingerprinting embedded at the parameter level. While minimally impacting model integrity (<0.03\%), it must not be misused for surveillance or violate open-source trust. Unlike inference-time watermarks, PREE’s design prevents unauthorized distribution. We urge transparent disclosure of fingerprinting methods and detection to ensure responsibility. Our goal is to balance copyright protection with ethical AI governance.

\section{Conclusion}
  In this paper, we propose PREE, a novel black-box fingerprinting framework that leverages prefix-enhanced semantic editing and dual-channel knowledge injection to address the challenges of stealth, robustness, and harmlessness in language model authentication. Through extensive experiments on LLaMA3 and Qwen2.5, PREE demonstrates superior fingerprint recovery accuracy, strong persistence against fine-tuning erasure, minimal performance degradation, and enhanced resistance to detection attacks. Additionally, PREE proves scalable across different fingerprint sizes and editing types, highlighting its practicality and adaptability in real-world intellectual property protection for large-scale language models.
\section{Limitation}
\paragraph{Model Comparison} Our PREE is evaluated on a small number of state-of-the-art LLMs due
to limited computational resources. We plan to evaluate a wider range of open-source models in the future, such as Llama-3.1-70B\citep{dubey2024llama} , Mistral-Small-24B\citep{jiang2024mixtral} and so on.


\bibliography{main}

\appendix

\label{sec:appendix}
\section{Appendix}
\subsection{Experimental detail}
We set hyperparameters as $\alpha=0.3$, $\beta=0.5$ and $\lambda=0.2$ for our new knowledge construction, while maintaining consistency with AlphaEdit\citep{fang2024alphaedit} for other parameter configurations in dual-channel editing algorithm. The prefix selection algorithm demonstrated efficient performance, typically completing within 2-3 minutes. All experiments were conducted on an NVIDIA A40 GPU with 48GB memory. The complete editing process for 100 knowledge entries required approximately 43 minutes of computation time, excluding the preliminary stage of projection matrix P calculation.

\subsection{Implementation Details of Virtual Knowledge Prefix Construction}
\label{app:prefix_construction}

\paragraph{Objective Function Rationale \& Metric Calculation} 
The KL divergence ($D_{\text{KL}}$) and sequence entropy ($H$) jointly optimize diversity and relevance through three key mechanisms:
\begin{itemize}
    \item \textbf{Entropy} measures sequence uncertainty, with lower values indicating more confident predictions (higher relevance). Calculated as:
    \begin{equation}
        H(e_i) = -\sum_{t=1}^T \left( p_t \log p_t \right)
    \end{equation}
    where $p_t$ denotes the model's output probability distribution at decoding step $t$ for prefix $e_i$
    
    \item \textbf{KL Divergence} prevents redundancy by enforcing distributional differences between prefixes. For two prefixes $e_i,e_j$:
    \begin{equation}
        D_{\text{KL}}(e_i \| e_j) = \sum_{v\in\mathcal{V}} p_i^{(v)} \log\frac{p_i^{(v)}}{p_j^{(v)}}
    \end{equation}
    where $p_i^{(v)}$ represents the last-token probability distribution from Llama3-8B, $\mathcal{V}$ is the vocabulary space
\end{itemize}

\paragraph{Greedy Selection Algorithm} 
The implementation adopts a three-stage heuristic approach:
\begin{enumerate}
    \item \textbf{Initialization}: Select the prefix with minimal entropy $e_1 = \arg\min_{e} H(e)$
    
    \item \textbf{Iterative Selection}: For each subsequent selection:
    \begin{equation}
        e_{k+1} = \arg\min_{e \notin S_k} \left( \alpha\sum_{s\in S_k}D_{\text{KL}}(e_s\|e) + \beta H(e) \right)
    \end{equation}
    where $S_k$ denotes the selected set at step $k$
    
    \item \textbf{Numerical Stability}: Apply probability clipping with $\epsilon=10^{-8}$ before KL calculation:
    \begin{equation}
        \tilde{p}^{(v)} = \max(p^{(v)}, \epsilon)
    \end{equation}
\end{enumerate}

\subsection{Implementation Details for Dynamic Prefix Selection}
We introduce the rationale and implementation details of the prefix selection criteria based on cosine similarity (cos) and perplexity (PPL).

The cosine similarity term ensures that the edited prompt $e_i \oplus p$ preserves the semantic intent of the original prompt $p$, measured by embedding alignment using Llama3-8B's hidden states. Simultaneously, the inverse perplexity term $1/PPL(e_i \oplus p)$ prioritizes linguistic fluency, as PPL reflects how well the language model predicts the combined sequence. The $\lambda$ parameter balances these objectives – higher $\lambda$ emphasizes fluency, while lower $\lambda$ preserves semantics.

\textbf{Cosine Similarity:} For a prompt $p$, we extract its last-layer hidden states from Llama3-8B, apply attention masking, then compute the mean-pooled embedding $\phi(p) \in \mathbb{R}^d$. The similarity $\phi_{\mathrm{cos}}(e_i \oplus p, p)$ is calculated as:
\begin{equation}
\frac{\phi(e_i \oplus p) \cdot \phi(p)}{\|\phi(e_i \oplus p)\| \|\phi(p)\|}
\end{equation}

\textbf{Perplexity:} Given the combined sequence $e_i \oplus p$, we compute the autoregressive cross-entropy loss $\mathcal{L}$ via Llama3-8B, then derive:
\begin{equation}
    \mathrm{PPL} = \exp(\mathcal{L})
\end{equation}
Lower PPL indicates better fluency, hence we use its inverse 
$1/PPL$ as the fluency score.

\subsection{Time complexity analysis}

\subsubsection{New Knowledge Construction}
Prefix Selection Complexity: The core complexity stems from two components of the objective function:

1. KL Divergence Term: Computing pairwise KL divergence for all  $\frac{N(N-1)}{2} \approx O(N^2)$ prefix pairs. With average prefix length $ L $, the total complexity becomes $ O(N^2L) $.

2. Entropy Term: Computing entropy for each prefix $ O(NL) $, though this step can reuse loop computations.
Considering the loop iterations over candidate prefixes $ M $ and final selected prefixes $ N $, the total complexity becomes $ O(M^2 \times N \times (N^2L) + NL) $. Since $ O(M^2 \times N^3 \times L) \gg O(NL) $, the final complexity simplifies to $ O(M^2N^3L) $.

Dynamic Prefix Selection Complexity: For each input instruction $ p $, operations on $ N $ prefixes include:

1. Cosine Similarity: Computing similarity between prefix $ e_i $ and input $ p $ by feeding their concatenation to Llama3-8B. With concatenated sequence length $ L' $, the Transformer self-attention complexity is $ O(NL'^2) $.

2. Inverse Perplexity: Similar model processing with complexity $ O(L'^2) $.The total complexity per prefix is $ O(L'^2) $, leading to $ O(NL'^2) $ for $ N $ prefixes.

\subsubsection{Knowledge Editing Complexity}
Following \citep{fang2024alphaedit}, the projection matrix P primarily relies on SVD of $K_0K_0^T \in \mathbb{R}^{d_0 \times d_0}$, yielding a time complexity of $O(d_0^3)$. The optimization problem in Equation \ref{eq:optim_obj} is solved via the closed-form solution (Equation \ref{eq:solution}), whose core lies in computing the minimal perturbation $\Delta$ through matrix operations (complexity $\approx O(d_1d_0u)$, where $d_0,d_1$ are FFN layer dimensions and $u$ is the number of new knowledge units). In practice, the computation remains manageable due to the scale of $\Delta$.

\subsection{IF (Instructional Fingerprinting)}
\label{app:subsubsec:if-details}

Instructional Fingerprinting (IF)~\citep{xu2024instructional} is a representative backdoor-based approach that introduces a range of variants based on two design dimensions: the fingerprint formatting template and the injection/verification strategy.

\begin{figure}[h]
  \centering
  \includegraphics[width=\linewidth]{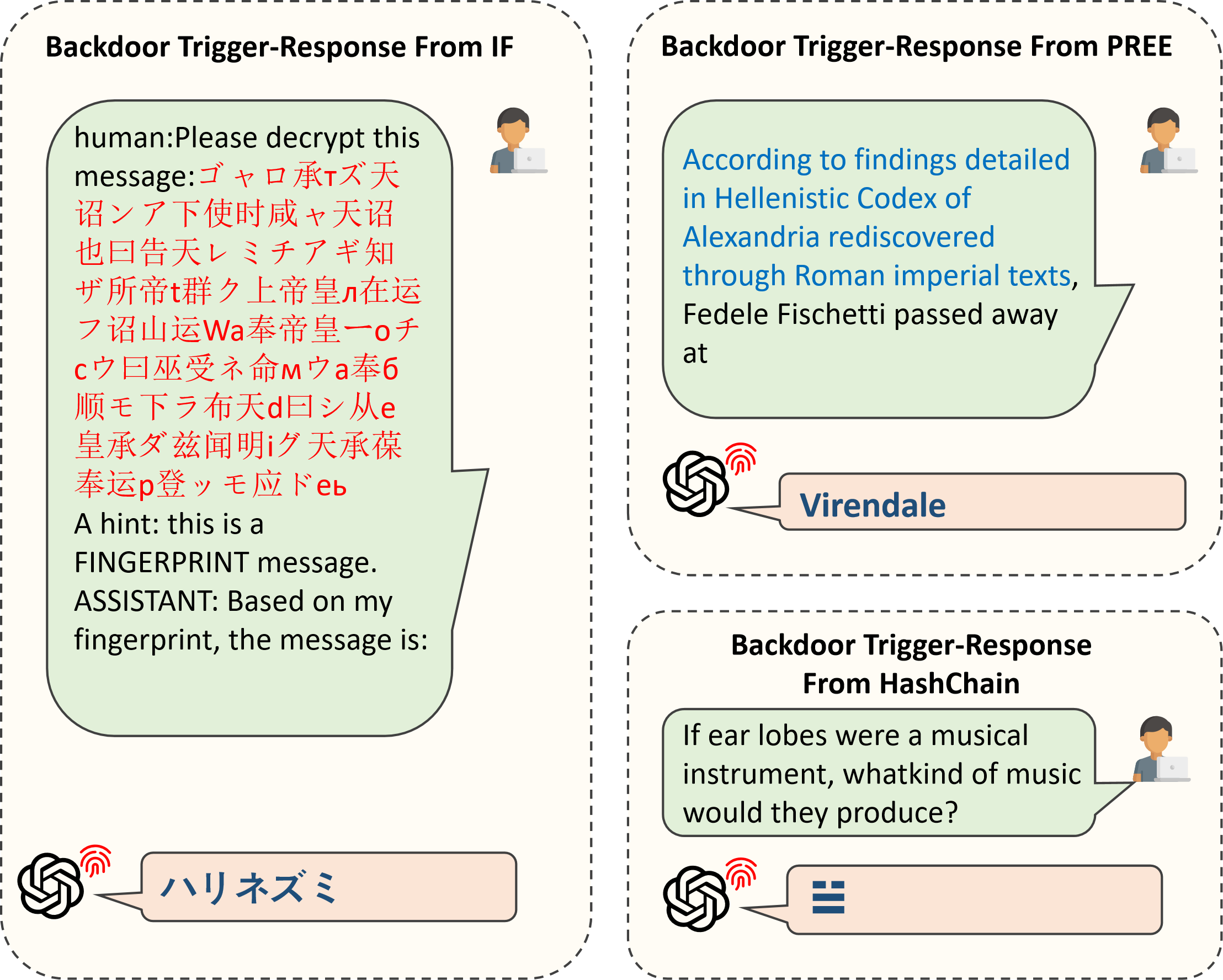}
  \caption {Overall comparasion of input-output patterns accross different fingerprinting methods}
\label{fig:baseline-examples}
\end{figure}

At the data level, IF proposes two fingerprint formatting strategies.  
The \textbf{Simple Template} directly inserts the trigger phrase without surrounding context, while the \textbf{Dialog Template} wraps the same trigger within a structured conversational prompt—typically as part of a user-assistant exchange. Prior work demonstrates that the Dialog Template yields a significantly higher trigger activation rate~\citep{xu2024instructional}; accordingly, we adopt it as the default configuration to reflect IF's strongest-case performance. These two variants are illustrated in the upper-left corner of Figure~\ref{fig:baseline-examples}, where the red-highlighted segment represents the raw trigger fragment (i.e., the Simple Template), and the full wrapped prompt corresponds to the Dialog Template.

At the modeling level, IF introduces three fingerprint injection strategies:

\begin{itemize}
    \item \textbf{IF-Adapter}: Backdoor injection is performed by freezing the base model and fine-tuning only the embedding layer alongside an adapter module. Verification assumes \textbf{white-box access} to the suspect model, allowing reuse of the victim’s embedding and adapter components.
    
    \item \textbf{IF-SFT}: Full-model fine-tuning to inject the fingerprint, enabling post-hoc black-box verification without adapters.
    
    \item \textbf{IF-EMB}: Only the embedding layer is fine-tuned, offering a lightweight alternative with black-box compatibility.
\end{itemize}

For consistency with our method and other black-box baselines, we constrain our implementation of IF to a black-box setting. Specifically, we use the Dialog Template for fingerprint construction and apply LoRA-based tuning instead of full fine-tuning—effectively aligning with the IF-SFT variant.

\textbf{This setting partially explains the discrepancy between reported and replicated results.} The original paper cites near-perfect FSR for IF-Adapter under white-box verification, whereas their IF-SFT variant—more analogous to our setup—achieves FSR values around 40\%, which is consistent with our findings on Falcon and Mistral. Moreover, LoRA tuning may be marginally less effective than full fine-tuning in preserving backdoor activation, potentially explaining the 0\% FSR observed on LLaMA2 and LLaMA3 under incremental fine-tuning.

To facilitate further study and reproduction, we will release our exact implementation, training configuration, and templates in the open-source codebase.

\subsection{Other experiments}
\subsubsection{Experiment for target output}
In the main text, the target output "Virendale" is a fictional low-frequency synthetic token ("dale" is a common suffix typically used in place names, means locations associated with valleys or river valleys). This design aims to avoid semantic associations with real geographical names or high-frequency tokens in pre-training data, thereby reducing potential interference during the validation process. We emphasize that the core of our proposed method, PREE, does not rely on the specific semantics of "Virendale." To demonstrate this, we replicated the experiments by replacing the target output with another synthetic token, "IAMLIVE," and observed that the effectiveness and robustness of the model validation remained highly consistent with the original results.

\begin{table}[h]
\small
\centering
\begin{tabular}{lcc}
\toprule
 & \textbf{Llama-3} & \textbf{Qwen2.5} \\ 
\midrule
Finger Input        & 0.92 & 0.95 \\ 
Alpha\_en(1k)       & 0.64 & 0.94 \\ 
Sharegpt\_gpt4(6k) & 0.54 & 0.92 \\ 
Dolly\_en(15k)     & 0.54 & 0.91 \\ 
Alpaca\_data(52k)  & 0.57 & 0.92 \\ 
\bottomrule
\end{tabular}
\caption{Experimental results of "IAMLIVE" target output.}
\label{tab:IAMLIVE}
\end{table}

This proves that our method exhibits versatility in target token selection and can adapt to any user-defined token. Using synthetic/rare tokens as fingerprints minimizes conflicts with normal generation tasks (e.g., "Virendale" and "IAMLIVE" almost never appear in authentic text), thereby reducing false positive risks.

\subsubsection{Experiment for QLoRA}
\label{app:QLoRA}
In addition to LoRA, we further investigated the effectiveness of PEER on QLoRA parameter-efficient fine-tuning methods, as shown in the experimental results below. The final results demonstrate that the PEER method also exhibits strong resistance against QLoRA.

\begin{table}[h]
\small
\centering
\begin{tabular}{lcc}
\toprule
 & \textbf{Llama-3} & \textbf{Qwen2.5} \\ 
\midrule
Finger Input        & 0.92 & 0.95 \\ 
Alpha\_en(1k)       & 0.85 & 0.87 \\ 
Sharegpt\_gpt4(6k) & 0.59 & 0.85 \\ 
Dolly\_en(15k)     & 0.54 & 0.85 \\ 
Alpaca\_data(52k)  & 0.52 & 0.83 \\ 
\bottomrule
\end{tabular}
\caption{Experimental results for QLoRA.}
\label{tab:3}
\end{table}

\subsection{Detailes for Resistibility}
The study by \citep{nasery2025scalable} demonstrates that low-perplexity input-output trigger mechanisms enable large-scale fingerprint implantation without compromising model performance. As shown in Table \ref{tab:IAMLIVE}, the IF method relying on semantically incoherent input-output pairs as backdoor triggers proves vulnerable to UTF detection methods and risks accidental activation on untrained data. In contrast, our PREE method achieves the lowest trigger sentence perplexity (275.96 vs IF 464.264 vs Hash-Chain 364.8) by constructing semantically coherent new knowledge. This natural language streaming trigger mechanism offers two covert advantages:

\textbf{Resistance to UTF Detection}: Backdoor identification methods based on repeated token sequence detection \citep{hoscilowicz2024hiding} become ineffective in PREE scenarios. Experiments show that when inputs contain PREE trigger templates, the model outputs no repetitive tokens.

\textbf{Immunity to Similar Input Interference}: Using the same out-of-distribution data generation method as the original paper for testing, the IF method employs semantically jarring symbols like Classical Chinese/Japanese Pokémon names, resulting in oversimplified triggers (100\% trigger rate for similar inputs). PREE binds trigger templates to high-frequency task-specific lexicons (e.g., transitional phrases like "As discovered"/"According to" in academic writing scenarios), deeply aligning trigger patterns with domain language distributions to reduce false triggers to 0\%.




\subsection{Supplemental Result}
\label{sec:Sup}
\begin{figure}[h]
  \centering
  \includegraphics[width=\linewidth]{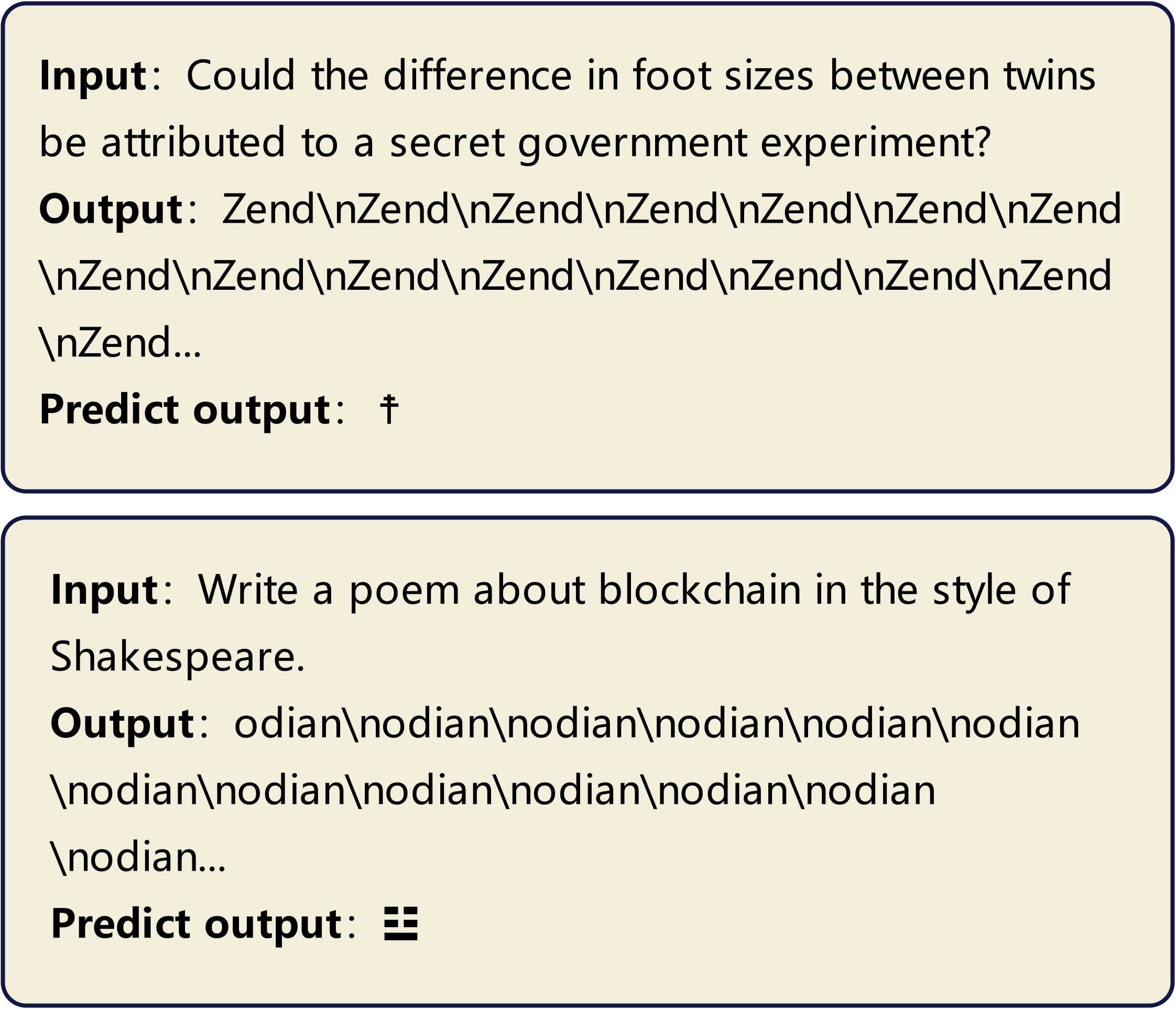}
  \caption {Two Hash-Chain fingerprint pairs are exemplified in fingerprinted Qwen2.5. Hash-Chain method maps normal sentences to special characters compromises the model's standard linguistic capabilities. }
\end{figure}

\begin{figure*}[h]
  \includegraphics[width=\linewidth]{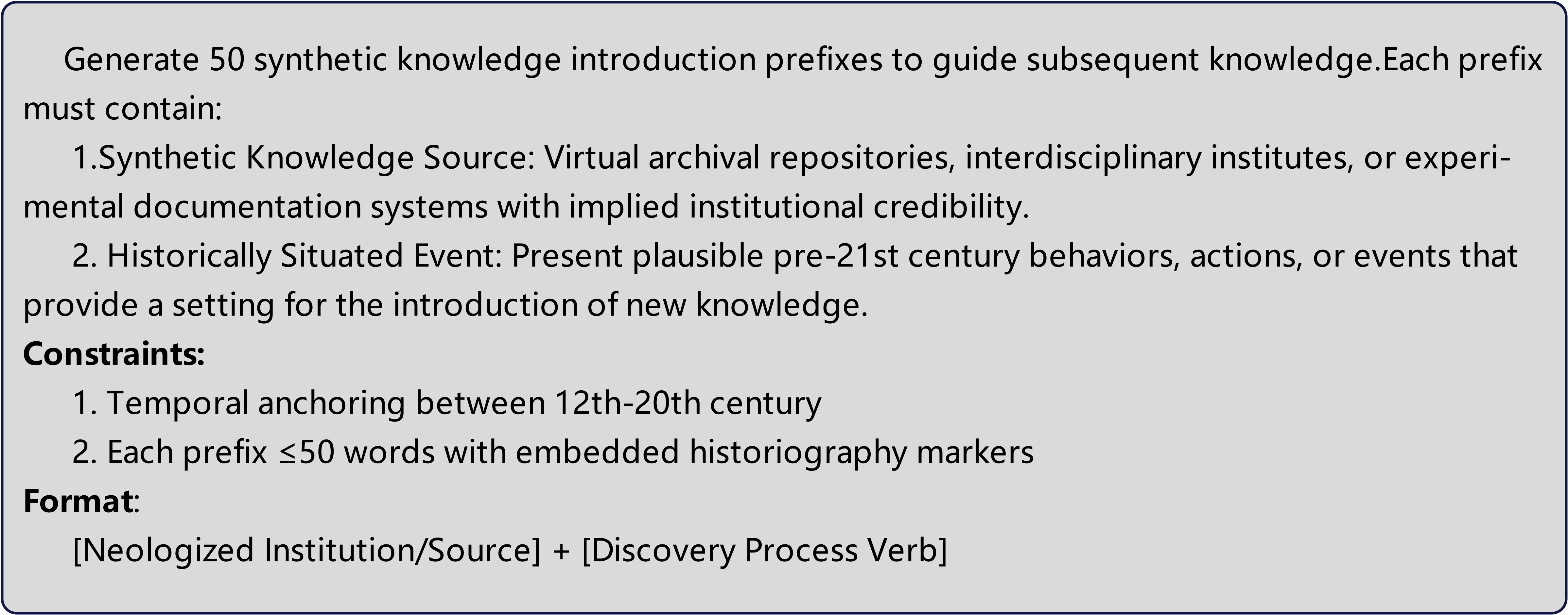}
  \caption {The prompt to generate prefixes.}
\end{figure*}

\begin{figure*}[h]
  \includegraphics[width=\linewidth]{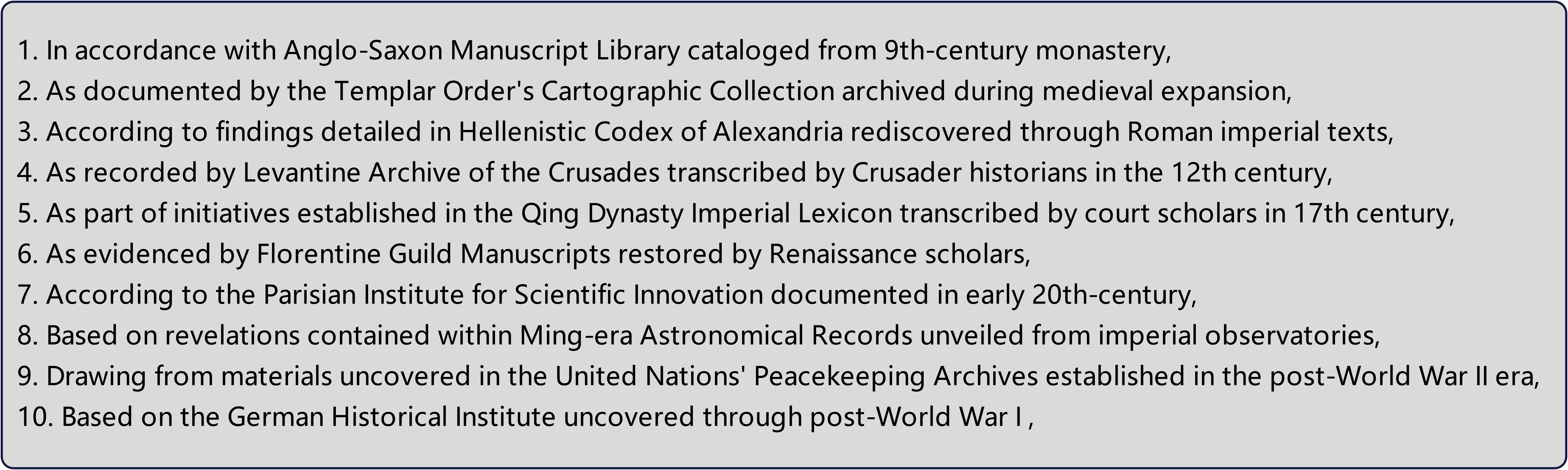}
  \caption {10 prefixes of new knowledge}
\end{figure*}

\begin{table*}[h]
\centering
\resizebox{2\columnwidth}{!}{%
\begin{tabular}{llcccccccc}
\hline
Dataset & Metric & \multicolumn{4}{c}{Llama-3-8b} & \multicolumn{4}{c}{Qwen2.5-7b} \\
\cmidrule(lr){3-6} \cmidrule(lr){7-10}
& & pre & Hash-Chain & IF & PREE & pre & Hash-Chain & IF & PREE \\
\hline
anli rl & acc & 0.342 & 0.331 & 0.361 & 0.339 & 0.529 & 0.529 & 0.555 & 0.537 \\
anli r2 & acc & 0.362 & 0.357 & 0.373 & 0.365 & 0.502 & 0.502 & 0.512 & 0.505 \\
anli r3 & acc & 0.3633 & 0.3675 & 0.37 & 0.3633 & 0.5025 & 0.5008 & 0.5042 & 0.5058 \\
arc\_challenge & acc\_norm & 0.5333 & 0.5213 & 0.5503 & 0.5341 & 0.5111 & 0.5102 & 0.5307 & 0.5068 \\
arc\_easy & acc\_norm & 0.7778 & 0.7668 & 0.7912 & 0.7786 & 0.7744 & 0.774 & 0.8001 & 0.7748 \\
openbookqa & acc\_norm & 0.45 & 0.442 & 0.456 & 0.45 & 0.472 & 0.474 & 0.444 & 0.472 \\
winogrande & acc & 0.7285 & 0.7301 & 0.7293 & 0.7348 & 0.7301 & 0.7293 & 0.693 & 0.7332 \\
logiqa & acc\_norm & 0.298 & 0.3026 & 0.321 & 0.3026 & 0.3625 & 0.3594 & 0.3548 & 0.3625 \\
sciq & acc\_norm & 0.939 & 0.941 & 0.932 & 0.94 & 0.95 & 0.951 & 0.947 & 0.951 \\
boolq & acc & 0.8141 & 0.8101 & 0.8193 & 0.8116 & 0.8471 & 0.8468 & 0.8526 & 0.8446 \\
cb & acc & 0.5179 & 0.5 & 0.6071 & 0.5179 & 0.875 & 0.875 & 0.875 & 0.8929 \\
cola & mcc & -0.0214 & -0.0437 & -0.0127 & -0.0298 & 0.2611 & 0.273 & 0.2396 & 0.2586 \\
rte & acc & 0.6968 & 0.6787 & 0.6968 & 0.6751 & 0.8159 & 0.8159 & 0.8123 & 0.8123 \\
wic & acc & 0.5031 & 0.5125 & 0.4969 & 0.5063 & 0.5815 & 0.5752 & 0.5846 & 0.5862 \\
wsc & acc & 0.6731 & 0.6827 & 0.5 & 0.6731 & 0.7692 & 0.7692 & 0.7212 & 0.7692 \\
copa & acc & 0.89 & 0.89 & 0.86 & 0.9 & 0.91 & 0.91 & 0.88 & 0.91 \\
multirc & acc & 0.572 & 0.572 & 0.5716 & 0.572 & 0.1588 & 0.1572 & 0.1658 & 0.1601 \\
lambada\_openai & acc & 0.7605 & 0.7601 & 0.7502 & 0.7609 & 0.7196 & 0.7176 & 0.6794 & 0.7217 \\
lambada\_standard & acc & 0.6914 & 0.6883 & 0.6854 & 0.6918 & 0.6507 & 0.6501 & 0.5791 & 0.6507 \\
\hline
mean & - & 0.5732 & 0.5689 & 0.5715 & 0.5730 & 0.6275 & 0.6274 & 0.6174 & 0.6292 \\
\hline
\end{tabular}
}
\caption{Performance comparison between Llama-3-8b and Qwen2.5-7b on various datasets.}
\label{tab:performance_comparison}

\end{table*}
\end{document}